# Gallium Substituted "114" $YBaFe_4O_7$: From a ferrimagnetic cluster glass to a cationic disordered spin glass

Tapati Sarkar*, V. Caignaert, V. Pralong and B. Raveau

*Laboratoire CRISMAT, UMR 6508 CNRS ENSICAEN,*

*6 bd Maréchal Juin, 14050 CAEN, France*

**Abstract**

The study of the ferrites $YBaFe_{4-x}Ga_xO_7$ shows that the substitution of Ga for Fe in $YBaFe_4O_7$ stabilizes the hexagonal symmetry for $0.40 \leq x \leq 0.70$, at the expense of the cubic one. Using combined measurements of a. c. and d. c. magnetization, we establish that Ga substitution for Fe in $YBaFe_4O_7$ leads to an evolution from a geometrically frustrated spin glass (for x = 0) to a cationic disorder induced spin glass (x = 0.70). We also find an intermediate narrow range of doping where the samples are clearly phase separated having small ferrimagnetic clusters embedded in a spin glass matrix. The origin of the ferrimagnetic clusters lies in the change in symmetry of the samples from cubic to hexagonal (and a consequent lifting of the geometrical frustration) as a result of Ga doping. We also show the presence of exchange bias and domain wall pinning in these samples. The cause of both these effects can be traced back to the inherent phase separation present in the samples.

* Corresponding author: Dr. Tapati Sarkar

e-mail:tapati.sarkar@ensicaen.fr

Fax: +33 2 31 95 16 00

Tel:  +33 2 31 45 26 32

**Introduction**

The recent studies of the "114" cobaltites $(Ln,Ca)_1BaCo_4O_7$ [1 – 5] and ferrites $(Ln,Ca)_1BaFe_4O_7$ [6 – 8] have generated a lot of interest in the scientific community because of their complex magnetic, electronic and thermoelectric properties [9]. These cobaltites and ferrites have the same basic structure, and are closely related to spinels and barium hexaferrites by their close packing of "$O_4$" and "$BaO_3$" layers. This close packing forms a 3-dimensional framework $[Fe_4O_7]_\infty$ (or $[Co_4O_7]_\infty$) consisting of corner-sharing $FeO_4$ (or $CoO_4$) tetrahedra, with the lanthanide elements occupying the octahedral sites of this framework. The triangular geometry of the cobalt (or iron) sublattices (**Fig. 1**) plays a dominant role in their magnetic properties. It was indeed shown that for hexagonal $LnBaCo_4O_7$ cobaltites (**Fig. 1 (a)**), there exists a strong competition between the 1 D magnetic ordering along the $\vec{c}$ direction in the "$Co_5$" trigonal bipyramids, and the magnetic frustration in the (001) plane built up of "$Co_3$" triangles [10, 11]. In fact, the magnetic frustration can be lifted by an orthorhombic distortion of the structure. This is illustrated by the concomitant structural and magnetic transitions that appear at low temperature in these cobaltites [1, 2, 12], and by the ferrimagnetic structure of $CaBaCo_4O_7$ [13]. Similarly, the "114" ferrites exhibit a competition between 1 D magnetic ordering and 2 D magnetic frustration, as has been shown for the hexagonal phases $CaBaFe_4O_7$ [6], and for $CaBaFe_{4-x}Li_xO_7$ [14]. But importantly, the "114" ferrites differ from the "114" cobaltites by the fact that the $LnBaFe_4O_7$ oxides exhibit a cubic structure [7]. Though the latter is closely related to the hexagonal structure, the iron sublattice is very different (**Fig. 1 (b)**), consisting of "$Fe_4$" tetrahedra instead of "$Fe_5$" bipyramids and "$Fe_3$" triangles. No structural transition appears at low temperature, and consequently, the cubic ferrites exhibit a spin glass behaviour due to a perfect geometrical frustration. Further, the $LnBaFe_4O_7$ series exhibits an oxidation state disorder. Unlike the case of $CaBaCo_4O_7$ [13], no charge ordering is observed in $LnBaFe_4O_7$, and this disorder is also important for the observed glassiness.

Recently, we showed that the substitution of a divalent cation, $Zn^{2+}$, for iron in $YBaFe_4O_7$, allowed the hexagonal symmetry to be stabilized at the detriment of the cubic one [15]. Paradoxically, it was observed that the substitution of this diamagnetic cation for $Fe^{2+}$ induces ferrimagnetism, in contrast to the spin glass behaviour of the undoped phase $YBaFe_4O_7$. In fact, a competition between ferrimagnetism and magnetic frustration was observed for the hexagonal phase $YBaFe_{4-x}Zn_xO_7$. This was interpreted as the effect of two antagonist phenomena: the partial lifting of the geometrical frustration due to the appearance of the hexagonal symmetry inducing a 1 D magnetic ordering, and the existence of cationic disordering favouring the glassy state.

Bearing in mind that the $Fe^{2+}:Fe^{3+}$ ratio is a crucial factor governing the magnetic properties of iron oxides, it must be emphasized that the substitution of $Zn^{2+}$ for $Fe^{2+}$ increases the average valence of iron, i.e. the $Fe^{2+}:Fe^{3+}$ ratio decreases from 3 in the spin glass phase $YBaFe_4O_7$ to 2.6 – 1.5 in the solid solution $YBaFe_{4-x}Zn_xO_7$ when x changes from 0.4 to 1.5 [15]. In order to further understand the role of the average valence of iron in the magnetic properties of these ferrites, we have investigated the possibility of substitution of a diamagnetic cation such as gallium for $Fe^{3+}$ in the $YBaFe_4O_7$ structure. In the present study of the ferrite $YBaFe_{4-x}Ga_xO_7$, we show that the introduction of gallium in the structure stabilizes the hexagonal symmetry, similar to the zinc substitution, but differently from the latter, the lifting of the geometrical frustration induces the formation of ferrimagnetic clusters embedded in a spin glass matrix, which tend to disappear as the gallium content increases, leading to a pure spin glass for higher Ga content, with a higher $T_g$ compared to $YBaFe_4O_7$.

**Experimental**

Phase-pure samples of $YBaFe_{4-x}Ga_xO_7$ [x = 0.40 – 0.70] were prepared by solid state reaction technique. The precursors used were $Y_2O_3$, $BaFe_2O_4$, $Ga_2O_3$, $Fe_2O_3$ and metallic Fe powder. First, the precursor $BaFe_2O_4$ was prepared from a stoichiometric mixture of $BaCO_3$ and $Fe_2O_3$ annealed at 1200°C for 12 hrs in air. In a second step, a stoichiometric mixture of $Y_2O_3$, $BaFe_2O_4$, $Ga_2O_3$, $Fe_2O_3$ and metallic Fe powder was intimately ground and pressed in the form of rectangular bars. The bars were then kept in an alumina finger, sealed in silica tubes under vacuum and annealed at 1100°C for 12 hrs. Finally, the samples were quenched to room temperature in order to stabilize the "114" phase.

The X-ray diffraction patterns were registered with a Panalytical X'Pert Pro diffractometer under a continuous scanning mode in the 2θ range 10° - 120° and step size Δ2θ = 0.017°. The cationic composition was confirmed by means of Energy Dispersive X-Ray Spectroscopy (EDS) technique using a Scanning Electron Microscope (ZEISS Supra 55). The d. c. magnetization measurements were performed using a superconducting quantum interference device (SQUID) magnetometer with variable temperature cryostat (Quantum Design, San Diego, USA). The a. c. susceptibility, $\chi_{ac}(T)$ was measured with a PPMS from Quantum Design with the frequency ranging from 10 Hz to 10 kHz. $H_{ac}$ was kept fixed at 10

Oe, while $H_{dc}$ was varied from 0 Oe to 2000 Oe. All the magnetic properties were registered on dense ceramic bars of dimensions ~ $4 \times 2 \times 2$ mm$^3$.

**Results and discussion**

Similar to Zn substitution, Ga substitution also favours the formation of the hexagonal phase at the expense of the cubic one. Nevertheless, the homogeneity range of the hexagonal YBaFe$_{4-x}$Ga$_x$O$_7$ solid solution is significantly different ($0.40 \leq x \leq 0.70$) vis – à – vis that of YBaFe$_{4-x}$Zn$_x$O$_7$ [15]. The cubic symmetry of YBaFe$_4$O$_7$ is retained for $0 \leq x \leq 0.20$, whereas the domain $0.20 < x < 0.40$ is biphasic, corresponding to a mixture of the cubic and hexagonal phases. On the other hand, for $x > 0.70$, several impurity phases appear, namely Y$_2$O$_3$ and Ga$_2$O$_3$. The cationic composition of the single phase obtained for the range $0.40 \leq x \leq 0.70$ using EDS analysis are also shown in **Table 1**.

*Structural characterization*

In **Fig. 2**, we show the X-ray diffraction (XRD) patterns of the two end members, (a) YBaFe$_{3.6}$Ga$_{0.4}$O$_7$ and (b) YBaFe$_{3.3}$Ga$_{0.7}$O$_7$ as representative examples. As stated before, the samples are seen to stabilize in the hexagonal symmetry with the space group *P6$_3$mc*. The Rietveld analysis of the lattice structure was done using the FULLPROF refinement program [16] and the fits are also shown in **Fig. 2**. All the samples in the range x = 0.40 – 0.70 were seen to stabilize in the same hexagonal symmetry.
The extracted cell parameters have been tabulated in **Table 1**. The ionic radius of Fe$^{3+}$ (0.49 Å) is very similar to that of Ga$^{3+}$ (0.47 Å). As can be seen from the extracted cell parameters shown in Table 1, *a* increases very slightly as x increases (an increase of only ~ 0.08 % as x increases from 0.4 to 0.7), while *c* shows a slight decrease (~ 0.11 %). This causes the cell volume to remain practically unchanged as a function of doping in accordance with the similar ionic radii of Fe$^{3+}$ and Ga$^{3+}$.

*D. C. magnetization studies*

In the "114" ferrites, it has been established earlier [8, 15] that ferrimagnetism is inherently linked with the cross-over from cubic to hexagonal symmetry. The doping-induced transition to the hexagonal symmetry involves a partial lifting of the 3D geometrical frustration, which is the root cause of the appearance of ferrimagnetism. Thus, we restrict our discussion of the magnetic data to the YBaFe$_{4-x}$Ga$_x$O$_7$ samples exhibiting hexagonal symmetry ($0.4 \leq x \leq 0.7$). We note here that the cubic samples ($x < 0.2$) are spin glasses similar to the undoped YBaFe$_4$O$_7$, and will not be discussed further.

The temperature dependence of d. c. magnetic susceptibility ($\chi_{dc}$ = M/H) was registered according to the standard zero field cooled (ZFC) and field cooled (FC) procedures. A magnetic field of 0.3 T was applied during the measurements. The measurements were done in a temperature range of 5 K to 300 K. The $\chi_{ZFC}$(T) and $\chi_{FC}$(T) curves of all the samples are shown in **Fig. 3**.

Undoped YBaFe$_4$O$_7$ is a spin glass with $T_g$ = 50 K [7]. The ZFC $\chi_{dc}$ versus T curve for YBaFe$_4$O$_7$ shows a pure cusp-like shape [7], typical of canonical spin glasses. A close look at **Fig. 3** reveals that the YBaFe$_{4-x}$Ga$_x$O$_7$ series of samples shows two different kinds of low temperature M$_{ZFC}$(T) curves vis-à-vis the shape of the curves. While the $\chi_{dc}$(T) curve of the highest substituted sample (x = 0.70) is very similar to that of canonical spin glasses (with a peak at ~ 50 K and a gradual decrease of the magnetization value below 50 K), for the lowest doped sample (x = 0.40), there is a sharp drop in the susceptibility value below the temperature at which $\chi_{ZFC}$ reaches its maximum value (75 K). The susceptibility value drops sharply till ~ 50 K (marked by a black arrow in **Fig. 3 (a)**), below which the decrease in $\chi_{ZFC}$ is more gradual. We note here that the measuring field that we have chosen (0.3 T) is smaller than the coercive field of the YBaFe$_{3.6}$Ga$_{0.4}$O$_7$ sample at T = 5 K (data shown later in **Fig. 5**). Thus, it is quite possible that the sharp drop in the susceptibility value occurs at the temperature where the coercive field of the sample becomes smaller than 0.3 T. However, following this argument, we should have obtained similar sharp drops in the $\chi_{ZFC}$(T) curves for the x = 0.5 and x = 0.6 samples also, as the coercive fields of the x = 0.5 and x = 0.6 samples at T = 5 K are also larger than 0.3 T. Instead, it is observed that the sharp drop in the $\chi_{ZFC}$(T) curve seen in the x = 0.40 sample is reduced to small kinks in the $\chi_{ZFC}$(T) curves for the x = 0.50 and x = 0.60 samples. More importantly, a study of the temperature dependence of the coercive field (H$_C$) of the x = 0.4 sample (data not shown here) reveals that H$_C$ becomes smaller than 0.3 T at ~ 17 K (i.e. at a temperature much below 50 K). This suggests that this feature is not a simple effect of the coercive field, rather it may have a more complex origin.

Another possibility is that this sudden decrease in $\chi_{ZFC}$(T) is due to domain wall pinning effects, which has, in fact, been observed previously in manganites [17 – 19]. Due to pinning, the domains would not freely rotate below the pinning temperature unless a high enough external field is present to overcome the pinned state. Upon zero field cooling, the domains would be pinned into random orientations. When a low field is applied (0.3 T in this case), the pinning effect still dominates over the effect of the applied magnetic field, and the magnetization is lower than what would be expected in the absence of pinning. However, the pinned domain walls can be thermally activated by increasing the temperature. This could be the cause of the visible jump in the $\chi_{ZFC}(T)$ curve at the temperature where the pinning effects are overcome by temperature (~ 50 K). As can be seen in **Fig. 3**, the pinning effect gradually decreases as the doping concentration is increased (the sharp drop in the $\chi_{ZFC}(T)$ curve seen in the x = 0.40 sample is reduced to small kinks in the $\chi_{ZFC}(T)$ curves for the x = 0.50 and x = 0.60 samples, and completely vanishes for the x = 0.70 sample). This indicates that the domain wall pinning is more prominent for small doping and vanishes for higher doping. This is counter-intuitive if the pinning is thought to arise due to the presence of Ga in the lattice. Thus, the fact that the domain wall pinning decreases with an increase in the doping concentration leads us to believe that this pinning does not arise from the disorder in the system. Rather, it has a more complex origin, which we discuss later.

Before we proceed further, we perform some additional measurements to make sure that the sharp drop in the $\chi_{ZFC}(T)$ curve observed in the lowest doped sample is indeed due to domain wall effects, and not arising from some additional (antiferro) magnetic transition in the sample. Thus, we subject the x = 0.40 sample to a degaussing experiment [17], wherein the sample was initially cooled from 300 K down to 5 K in a zero external magnetic field. At 5 K, a large magnetic field (5 T) was applied. The magnetic field was then reduced to zero, and the sample was degaussed at 5 K by cycling a field of reducing intensity so that the remanent magnetization of the sample was reduced to zero. A magnetic field of 0.3 T was then applied, and the $\chi_{dc}(T)$ curve was recorded while warming the sample, in the same way as ZFC magnetization is recorded. The results are shown in **Fig. 4**. We find that the sudden sharp drop observed in the normally obtained ZFC curve (**Fig. 4 (a)**) vanishes when the sample is subjected to a high enough magnetic field, and then degaussed (**Fig. 4 (b)**). This experiment, thus, provides supplementary evidence that the sudden drop in magnetization seen below 75 K is not due to any kind of (antiferro) magnetic transition in the sample, but is probably associated with domain wall pinning effects. The high magnetic field (5 T) to which the sample was subjected was sufficient for domain wall displacements thereby destroying the pinning. In fact, a ZFC magnetization recorded under a high enough field of 5 T (see inset of **Fig. 4 (a)**) does not show any sharp drop in the magnetization of the sample.

The d. c. magnetization M(H) curves of all the samples registered at T = 5 K are shown in **Fig. 5**. The virgin curves of the M(H) loops are represented by black circles while the rest of the M(H) loops are shown by red lines. The first notable point is that for higher Ga substitution (x = 0.70), the M(H) loop is narrow and S – shaped (**Fig. 5 (d)**), which is quite typical of spin glasses and superparamagnets. On the other hand, for lower Ga substitution, the samples have larger loops with higher values of the coercivity and remanent magnetization, which keep increasing as the doping concentration is decreased. This indicates the presence of a higher degree of magnetic ordering in the lower doped samples as compared to the higher doped ones.

Another feature which strongly supports the presence of domain wall pinning is that the virgin curve of the x = 0.40 sample lies slightly outside the main M(H) loop (**Fig. 5 (a)**). This unusual feature of the virgin curve lying outside the main hysteresis loop has earlier been associated with irreversible domain wall motion in spinel oxides [20]. We also note that the virgin curve starts to shift inside the main M(H) loop as the doping concentration (x) is increased, and for the x = 0.70 sample, the entire virgin curve lies inside the main M(H) loop (**Fig. 5 (d)**). We again note that the domain wall pinning is more prominent in the samples with lower doping concentration.

In **Fig. 6**, we once again show the d. c. magnetization M(H) curves of all the samples registered at T = 5 K, but in three different modes: (i) normal ZFC mode, (ii) FC mode with a magnetic field of 2 T and (iii) FC mode with a magnetic field of H = - 2 T. In the ZFC mode, the samples were cooled from 300 K to 5 K in zero external magnetic field, following which M versus H curves were registered. In the FC mode, on the other hand, the samples were cooled from 300 K to 5 K in the presence of an external magnetic field (H = 2 T or – 2 T), and then M versus H curves were registered. For the highest substituted sample (x = 0.70), all three M(H) curves overlap each other (**Fig. 6 (d)**). However, for lower doping concentration, the field cooled M(H) loops exhibit shifts both in the field as well as in the magnetization axes. This is the exchange bias phenomenon [21, 22] that results from exchange interaction between ferromagnetic and antiferromagnetic materials. In our $YBaFe_{4-x}Ga_xO_7$ samples with low Ga concentration, the observed exchange bias can be explained in terms of interfacial exchange coupling between the coexisting ferrimagnetic cluster glass and the disordered spin glass-like phases. This exchange bias effect arising from the inherent phase separation in the $YBaFe_{4-x}Ga_xO_7$ samples is similar to that seen in some disordered manganites [23]. As can be seen from **Fig. 6**, the exchange bias effect keeps decreasing as the doping concentration is increased, and as stated before, it completely disappears for the doping concentration x = 0.70. We can explain this observation by considering that for lower Ga concentration, the samples consist of coexisting ferrimagnetic clusters embedded in a spin glass-like matrix, but as the doping concentration by the diamagnetic cation (Ga) is increased, the ferrimagnetic clusters are progressively reduced and we ultimately get a homogeneous spin glass (x = 0.70). The absence of phase separation in the x = 0.70 sample, thus, results in an absence of the exchange bias effect. The fact that the $YBaFe_{4-x}Ga_xO_7$ samples with lower Ga concentration are intrinsically phase separated, while the x = 0.70 sample is not, also affords us an alternative explanation for the domain wall pinning effects seen in the lower doped samples. As stated previously, the fact that the domain wall pinning is seen in the lower doped samples and not in the x = 0.70 sample means that it cannot arise from the disorder in the system. Rather, we believe that the pinning arises from an interplay between the two magnetic phases in the phase separated samples. Such a domain wall pinning process arising from the interplay between two coexisting magnetic phases has been seen earlier in intermetallic alloys [24]. We also note that apart from the exchange bias effect in the lower doped samples, field cooling also results in an overall increase in the coercivity and remanence magnetization values. This can be interpreted as an increase in the volume fraction of the magnetically ordered phase when the samples are cooled in the presence of an external magnetic field. Since field cooling improves the remanence in the lower doped samples, hence it was important to register M-H curves after field cooling with positive as well as negative cooling fields and check whether the M-H loops shift in opposite directions in order to confirm that there is indeed a genuine exchange bias effect in the lower doped samples.

*A. C. magnetic susceptibility studies*

The measurements of the a. c. magnetic susceptibility $\chi'_{ac}(T, f, H)$ were performed at different frequencies ranging from 10 Hz to 10 kHz, and different external magnetic fields ($H_{dc}$) ranging from 0 T to 0.2 T using a PPMS facility. The amplitude of the a. c. magnetic field was ~ 0.001 T In **Fig. 7** and **Fig. 8**, we show the temperature dependence of the real (in-phase) component of the a. c. susceptibility in the temperature range 10 K – 160 K of the lowest doped sample (x = 0.40) and the highest doped sample (x = 0.70) respectively, with a measuring frequency of 10 kHz and in zero magnetic field ($H_{dc}$ = 0).

From **Fig. 7**, it is clear that the $\chi'(T)$ curve of the x = 0.40 sample shows two features, one at T = 86 K (marked by a black arrow), and the second at T = 44 K (marked by a red arrow). While the high temperature feature at 86 K is a clear peak, the low temperature one at 44 K is a shoulder like feature and is more clearly evidenced in the imaginary (out-of-phase) component of the a. c. susceptibity (shown in inset (a) of **Fig. 7**). Repeating the measurements using four measuring frequencies (ranging from 10 Hz – 10 kHz) reveals that both the features are frequency dependent (this is shown in inset (b) of **Fig. 7**). The x = 0.70 sample, on the other hand, shows only one feature (at T = 60 K) in the $\chi'(T)$ and $\chi''(T)$ curves, shown in the main panel and inset (a) of **Fig. 8** respectively. Inset (b) of **Fig. 8** shows the $\chi'(T)$ curves measured using four different frequencies, and reveals that this peak at 60 K is also strongly frequency dependent. We note that the imaginary part of $\chi$ for $YBaFe_{3.6}Ga_{0.4}O_7$ is ~ 8 % of the real part of $\chi$. This is commonly found for systems where the spin domains are relatively large. $YBaFe_{3.6}Ga_{0.4}O_7$ can thus be described as a cluster glass. On the other hand, the imaginary part of $\chi$ for $YBaFe_{3.3}Ga_{0.7}O_7$ is significantly smaller (about 2.4% of the real part of $\chi$) which indicates that $YBaFe_{3.3}Ga_{0.7}O_7$ is closer to a canonical spin glass.

Although all the features in the a. c. susceptibility curves of the two samples described above have a single commonality, in that they are all strongly frequency dependent, but the nature and origin of these peaks can be quite different. Specifically, we need to establish the nature of the low temperature shoulder in the x = 0.4 sample at T ~ 50 K. Since it occurs close to the temperature where we have evidenced domain wall pinning from the d. c. magnetic data, it is tempting to attribute this low temperature shoulder in the a. c. susceptibility data to the same phenomenon. However, it is also possible that this low temperature feature is a superparamagnetic effect of the ferrimagnetic clusters. To investigate this, we perform further measurements of $\chi'(T)$ and $\chi''(T)$ of the two limiting samples (x = 0.40 and x = 0.70) in the presence of different external magnetic fields $H_{dc}$ ranging from 0 to 0.2 T. The results for the $YBaFe_{3.3}Ga_{0.7}O_7$ sample are shown in **Fig. 9**. It is seen that both $\chi'$ and $\chi''$ are strongly suppressed by the magnetic field. The peak also shows a continual shift towards lower temperature as the external magnetic field is increased (see the black arrows in **Fig. 9**). This is typical of the behaviour of a spin glass freezing temperature under the influence of magnetic field.

In **Fig. 10**, we show the results for the $YBaFe_{3.6}Ga_{0.4}O_7$ sample. It is seen that while the high temperature peak was significantly suppressed in the presence of external magnetic field, the low temperature peak was largely unaffected relative to the case $H_{dc} = 0$. This rules out the scenario of superparamagnetism being responsible for this low temperature feature, and confirms that the 50 K anomaly arises due to enhanced domain wall pinning, signatures of which have been observed and commented upon earlier in the d. c. magnetization measurements also. We also note that the high temperature peak shifts towards higher temperature as the external magnetic field is increased (see the black arrow in **Fig. 10 (a)**). This is not expected for a pure spin glass freezing transition, where the peak should shift towards lower temperature as the magnetic field is increased. We believe that this anomaly arises because the x = 0.40 sample is not a pure spin glass, rather it is a phase separated sample consisting of ferrimagnetic clusters embedded in a spin glass matrix.

**Conclusion**

These results show that the substitution of $Ga^{3+}$ for $Fe^{3+}$ in $YBaFe_4O_7$ induces a structural transition from cubic to hexagonal, similar to the substitution of $Zn^{2+}$ for $Fe^{2+}$ in this compound. Though the two types of substitutions induce a lifting of the geometrical frustration through a change of the structure, the effect of these diamagnetic cations upon the magnetic properties is different. A strong ferrimagnetic component is induced by zinc substitution [15], whereas Ga substitution leads to the formation of ferrimagnetic clusters embedded in a spin glass matrix, essentially leading to phase separation in the samples. The difference originates from the opposite evolution of the $Fe^{3+}$:$Fe^{2+}$ ratio as the substitution rate increases in the two cases. Both $Fe^{3+}$ and $Fe^{2+}$ exhibit the high spin configuration since they have a tetrahedral coordination in these ferrites. Thus, the magnetic moment induced by the $e_g^2 t_{2g}^3$ $Fe^{3+}$ cations should be much higher than that induced by the $e_g^3 t_{2g}^3$ $Fe^{2+}$ cations. Hence, an increase in the $Fe^{3+}$:$Fe^{2+}$ ratio should favour stronger magnetic interactions. In the case of $Zn^{2+}$ doping, the $Fe^{3+}$:$Fe^{2+}$ ratio increases, thereby favouring the appearance of ferrimagnetism. On the other hand, for $Ga^{3+}$ doping, the $Fe^{3+}$:$Fe^{2+}$ ratio decreases, thereby inducing only weak ferrimagnetism and cluster formation. In both series, $YBaFe_{4-x}Ga_xO_7$ and $YBaFe_{4-x}Zn_xO_7$, a dilution effect is observed with an increase in the doping concentration. As a consequence, ferrimagnetism is weakened for higher concentrations in the Zn – phase. In the Ga – phase, the ferrimagnetic clusters are magnetically coupled by exchange interactions mediated through the surrounding spin glass matrix. For higher Ga concentrations, the exchange coupling between the ferrimagnetic clusters becomes less efficient, ultimately leading to the formation of a pure spin glass phase for x = 0.70, which is similar to the pristine sample (x = 0), but with a slightly higher $T_g$ (60 K). The Ga – substituted phase also differs from the Zn – phase by the presence of exchange bias and domain wall pinning. The cause of both these effects can be traced back to the inherent phase separation present in the samples.

**Acknowledgements**

We acknowledge the CNRS and the Conseil Regional of Basse Normandie for financial support in the frame of Emergence Program and N°10P01391. V. P. acknowledges support by the ANR-09-JCJC-0017-01 (Ref: JC09_442369).

**Table captions**

**Table 1**: Cell parameters as obtained from the Rietveld refinement of X-ray powder diffraction data.

**Figure captions**

**Figure 1:** Schematic representation of (a) hexagonal $LnBaCo_4O_7$ and (b) cubic $YBaFe_4O_7$ (adapted from Ref. 8). For details, see text.

**Figure 2:** X-ray diffraction pattern along with the fits for (a) $YBaFe_{3.6}Ga_{0.4}O_7$ and (b) $YBaFe_{3.3}Ga_{0.7}O_7$.

**Figure 3:** Temperature dependence of the magnetic susceptibility ($\chi_{dc}$ = M/H) collected according to zero field cooling (ZFC) and field cooling (FC) processes for $YBaFe_{4-x}Ga_xO_7$ (a) x = 0.40, (b) x = 0.50, (c) x = 0.60 and (d) x = 0.70, measured at B = 0.3 T.

**Figure 4:** $\chi_{ZFC}(T)$ curves of $YBaFe_{3.6}Ga_{0.4}O_7$ recorded (a) in the ZFC mode without degaussing, and (b) after applying a magnetic field of 5 T and degaussing the ZFC sample (see text for details). The inset in (a) shows $\chi_{ZFC}(T)$ recorded under a magnetizing field of 5 T.

**Figure 5:** M (H) curves for $YBaFe_{4-x}Ga_xO_7$ (a) x = 0.40, (b) x = 0.50, (c) x = 0.60 and (d) x = 0.70, registered at T = 5 K. The virgin curves are shown in black circles, while the rest of the hysteresis loops are shown in red lines.

**Figure 6:** M (H) curves for $YBaFe_{4-x}Ga_xO_7$ (a) x = 0.40, (b) x = 0.50, (c) x = 0.60 and (d) x = 0.70, registered at T = 5 K, measured after zero field cooling (red open circles), field cooling in a field of 2 T (black lines) and field cooling in a field of - 2 T (blue lines).

**Figure 7:** Temperature dependence of the real (in-phase) component of a. c. susceptibility for $YBaFe_{3.6}Ga_{0.4}O_7$ as a function of temperature measured in zero magnetic field ($H_{dc}$ = 0), using a frequency of 10 kHz. Inset (a) shows the imaginary (out-of-phase) component of the a. c. susceptibility, and inset (b) shows the real (in-phase) component of the a. c. susceptibility measured using four different frequencies.

**Figure 8:** Temperature dependence of the real (in-phase) component of a. c. susceptibility for YBaFe$_{3.3}$Ga$_{0.7}$O$_7$ as a function of temperature measured in zero magnetic field (H$_{dc}$ = 0), using a frequency of 10 kHz. Inset (a) shows the imaginary (out-of-phase) component of the a. c. susceptibility, and inset (b) shows the real (in-phase) component of the a. c. susceptibility measured using four different frequencies.

**Figure 9:** The (a) real (in-phase) and (b) imaginary (out-of-phase) component of a. c. susceptibility for YBaFe$_{3.3}$Ga$_{0.7}$O$_7$ as a function of temperature. The driving frequency was fixed at $f$ = 1 kHz and H$_{ac}$ = 10 Oe. Each curve was obtained under different applied static magnetic field (H$_{dc}$) ranging from 0 T to 0.2 T.

**Figure 10:** The (a) real (in-phase) and (b) imaginary (out-of-phase) component of a. c. susceptibility for YBaFe$_{3.6}$Ga$_{0.4}$O$_7$ as a function of temperature. The driving frequency was fixed at $f$ = 1 kHz and H$_{ac}$ = 10 Oe. Each curve was obtained under different applied static magnetic field (H$_{dc}$) ranging from 0 T to 0.2 T.

**Table 1**

| Doping concentration (x) | Crystal system (Space group) | Unit cell parameters | | c / a | $\chi^2$ | x as obtained from EDS analysis |
| --- | --- | --- | --- | --- | --- | --- |
| | | a (Å) | c (Å) | | | |
| 0.40 | Hexagonal ($P6_3mc$) | 6.320 (1) | 10.383 (1) | 1.6428 | 3.02 | 0.43 (2) |
| 0.50 | Hexagonal ($P6_3mc$) | 6.322 (1) | 10.376 (1) | 1.6413 | 3.15 | 0.50 (1) |
| 0.60 | Hexagonal ($P6_3mc$) | 6.323 (1) | 10.374 (1) | 1.6407 | 2.94 | 0.59 (1) |
| 0.70 | Hexagonal ($P6_3mc$) | 6.325 (1) | 10.372 (1) | 1.6398 | 3.46 | 0.74 (3) |

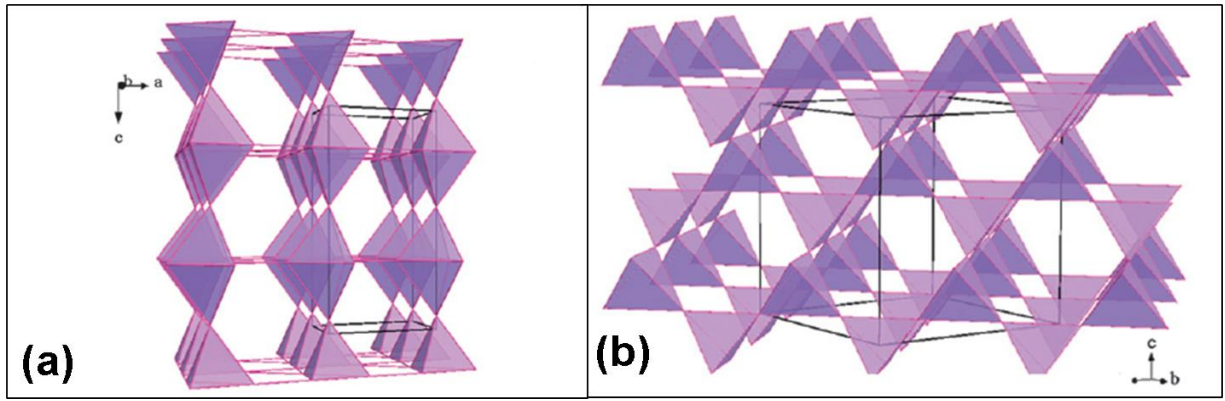

Fig. 1. Schematic representation of (a) hexagonal LnBaCo$_4$O$_7$ and (b) cubic YBaFe$_4$O$_7$ (adapted from Ref. 8). For details, see text.

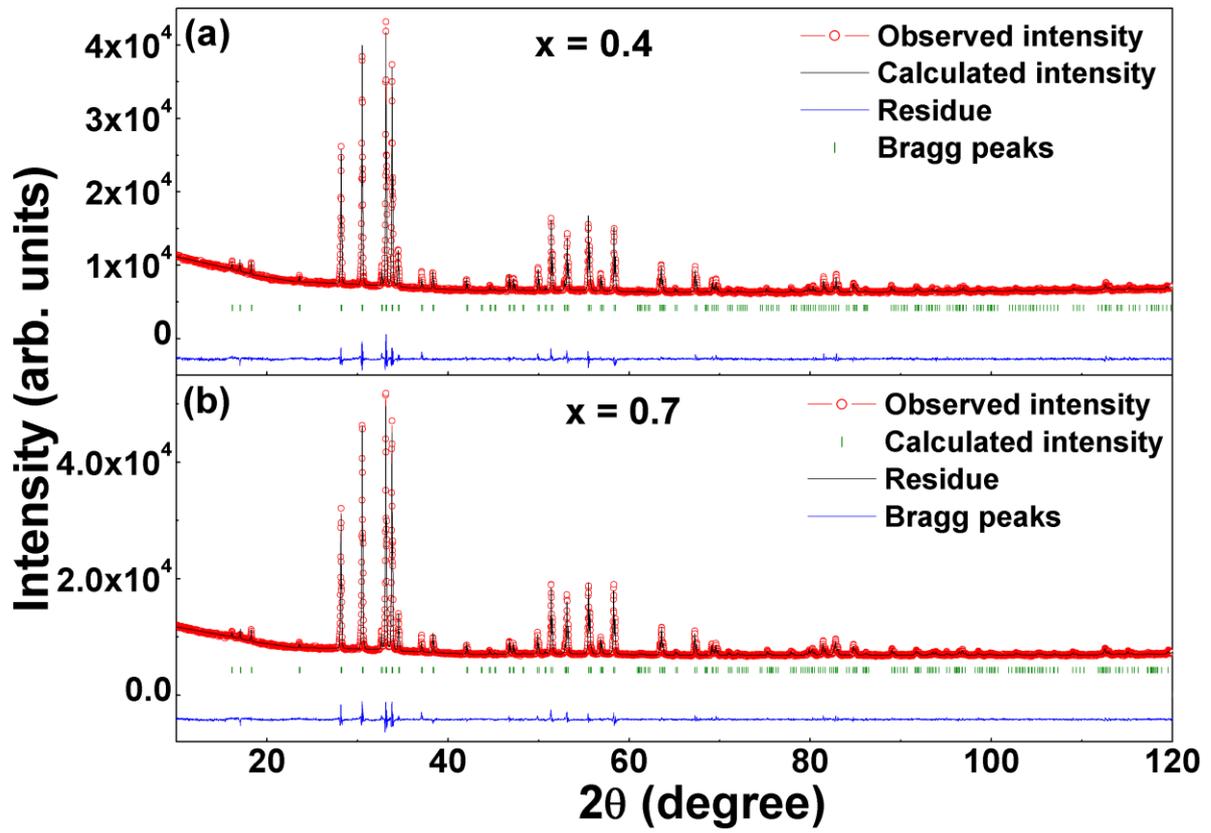

Fig. 2. X-ray diffraction pattern along with the fits for (a) $YBaFe_{3.6}Ga_{0.4}O_7$ and (b) $YBaFe_{3.3}Ga_{0.7}O_7$.

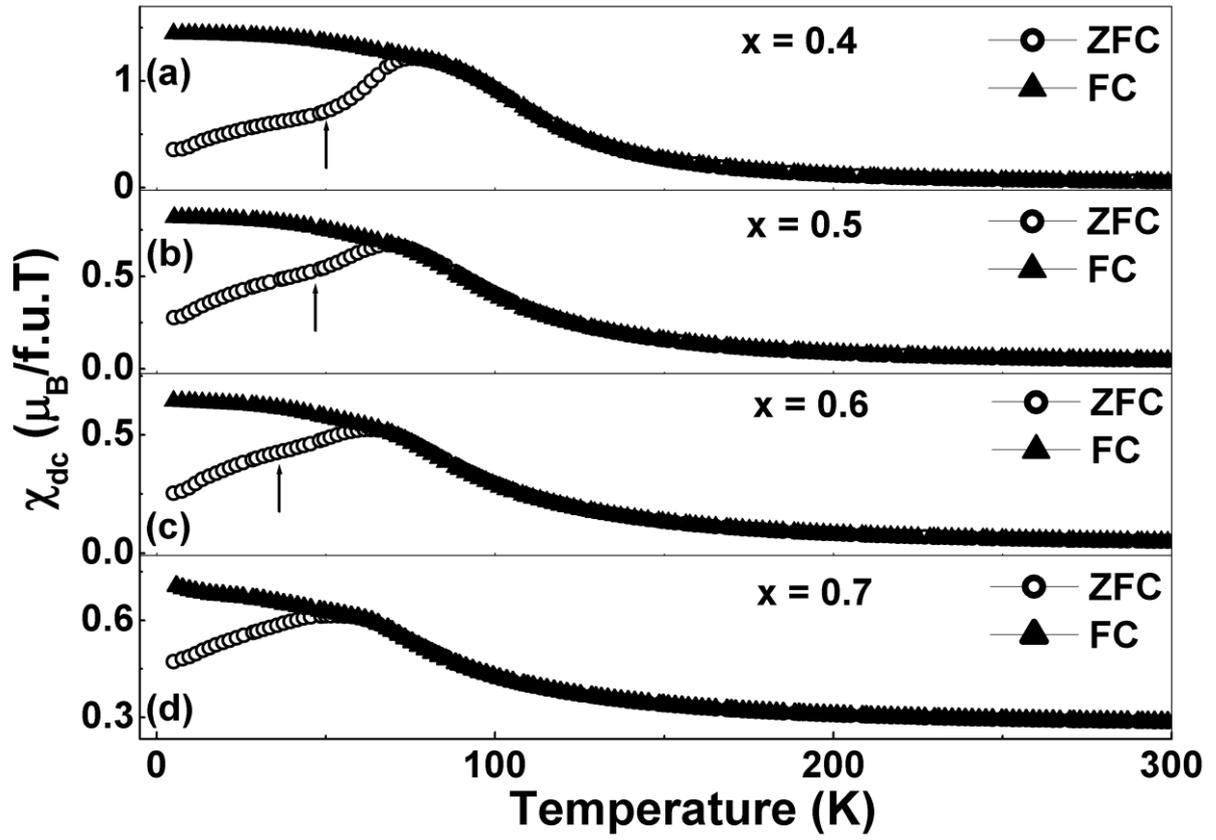

Fig. 3. Temperature dependence of the magnetic susceptibility ($\chi_{dc}$ = M/H) collected according to zero field cooling (ZFC) and field cooling (FC) processes for $YBaFe_{4-x}Ga_xO_7$ (a) x = 0.40, (b) x = 0.50, (c) x = 0.60 and (d) x = 0.70, measured at H = 0.3 T.

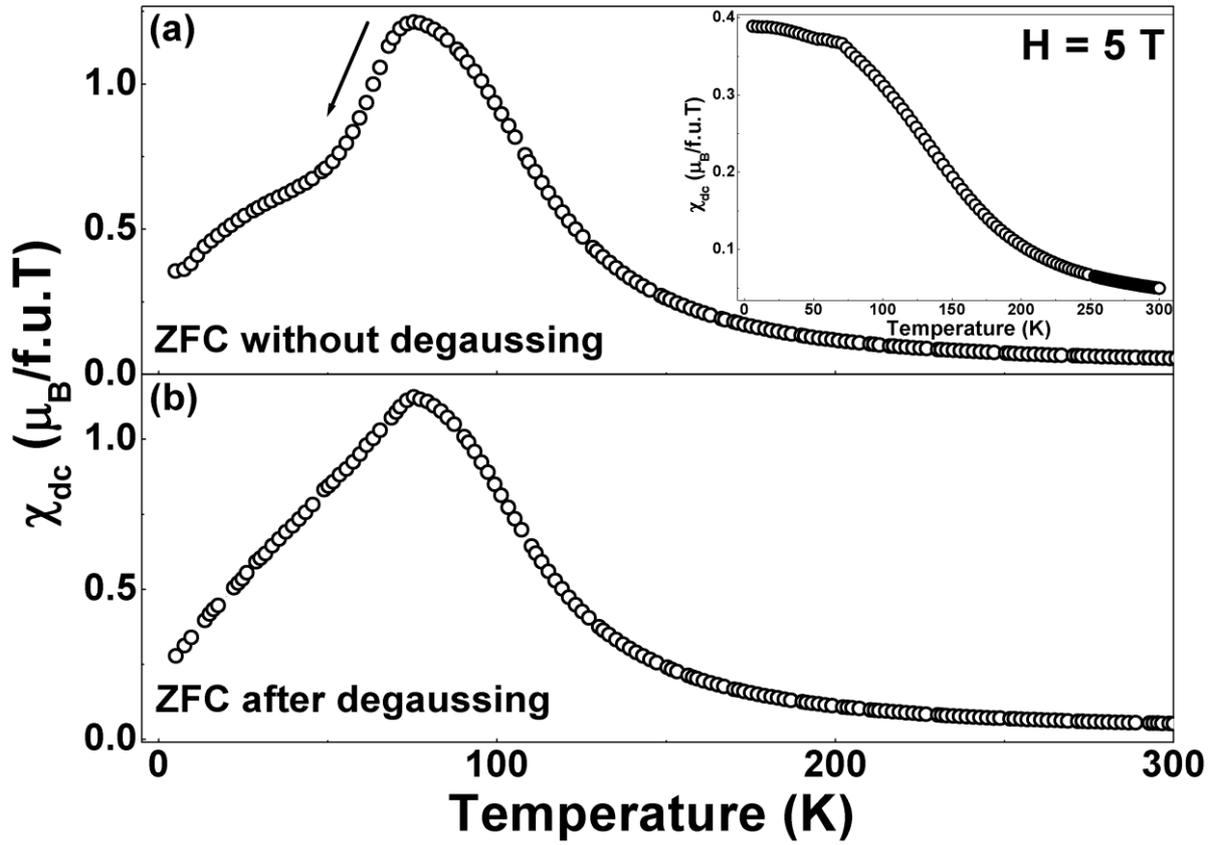

Fig. 4. $\chi_{ZFC}(T)$ curves of YBaFe$_{3.6}$Ga$_{0.4}$O$_7$ recorded (a) in the ZFC mode without degaussing, and (b) after applying a magnetic field of 5 T and degaussing the ZFC sample (see text for details). The inset in (a) shows $\chi_{ZFC}(T)$ recorded under a magnetizing field of 5 T.

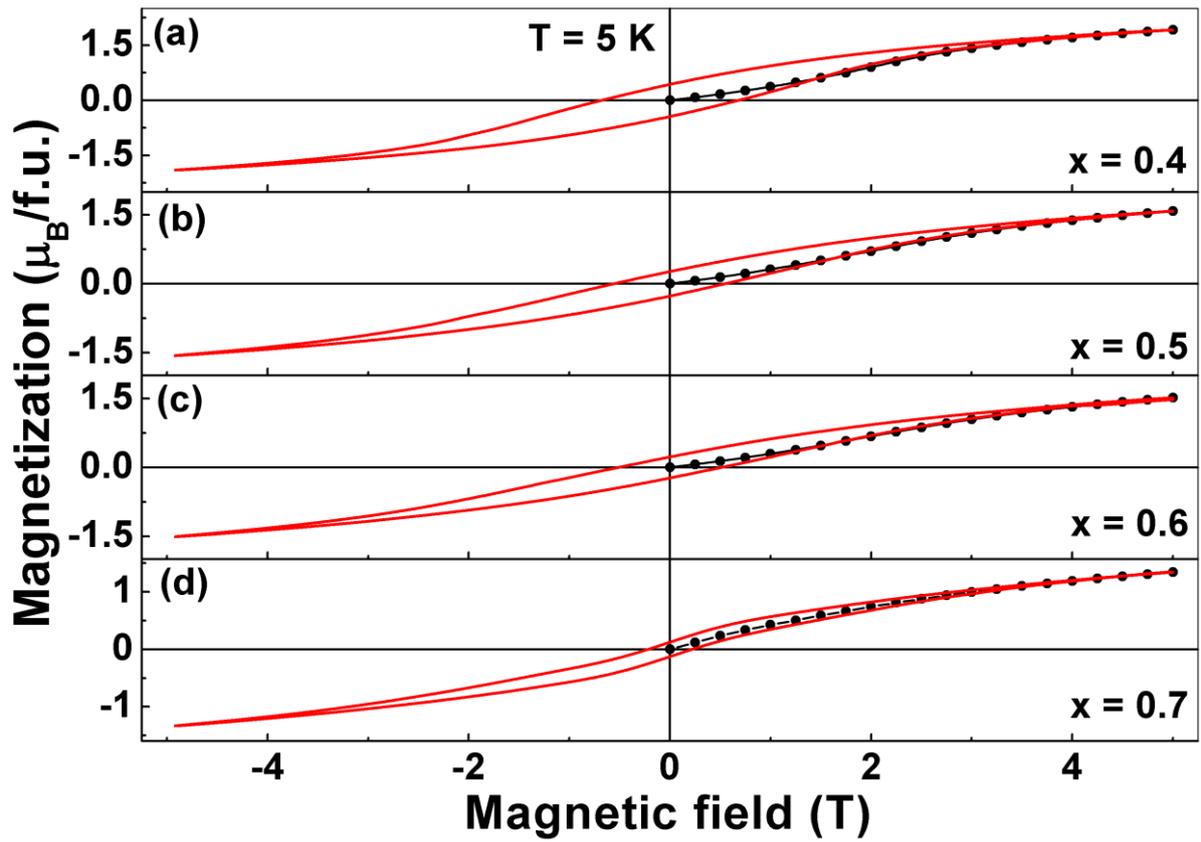

Fig. 5. M (H) curves for YBaFe$_{4-x}$Ga$_x$O$_7$ (a) x = 0.40, (b) x = 0.50, (c) x = 0.60 and (d) x = 0.70, registered at T = 5 K. The virgin curves are shown in black circles, while the rest of the hysteresis loops are shown in red lines.

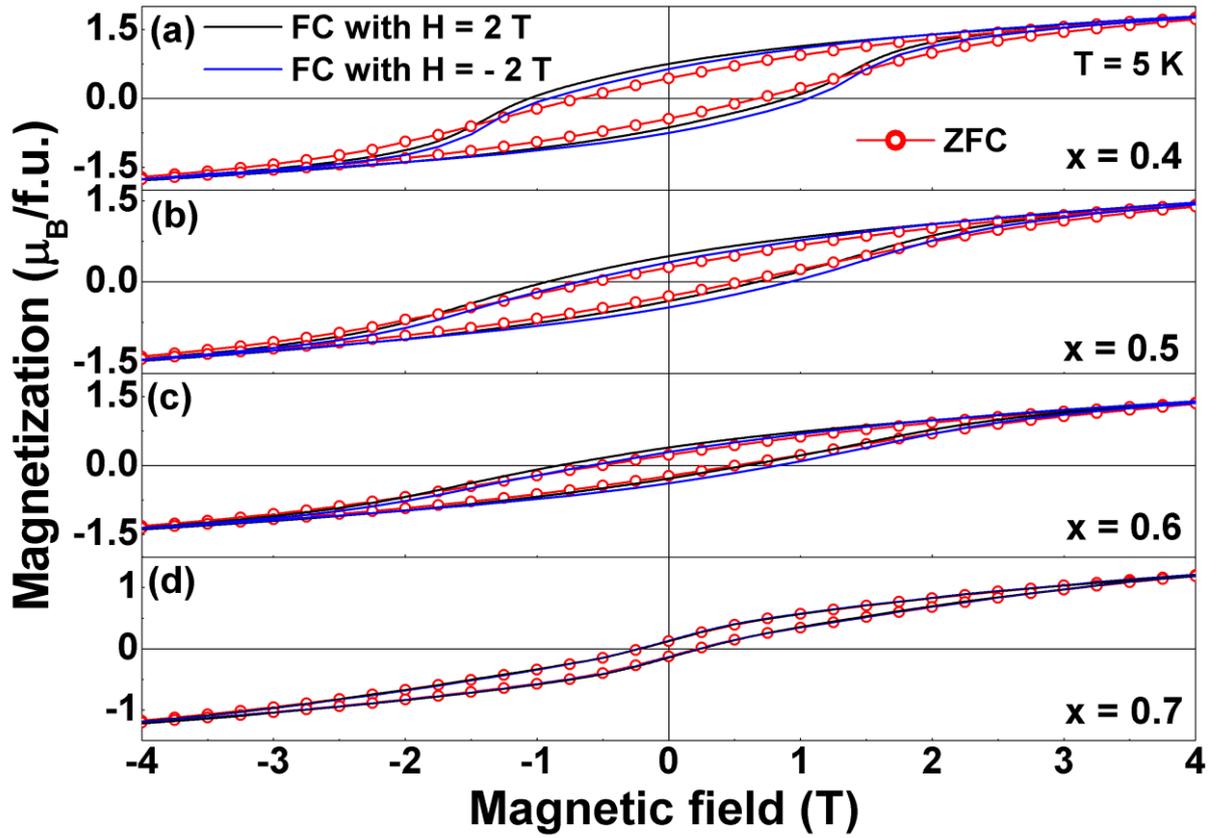

Fig. 6. M (H) curves for YBaFe$_{4-x}$Ga$_x$O$_7$ (a) x = 0.40, (b) x = 0.50, (c) x = 0.60 and (d) x = 0.70, registered at T = 5 K, measured after zero field cooling (red open circles), field cooling in a field of 2 T (black lines) and field cooling in a field of - 2 T (blue lines).

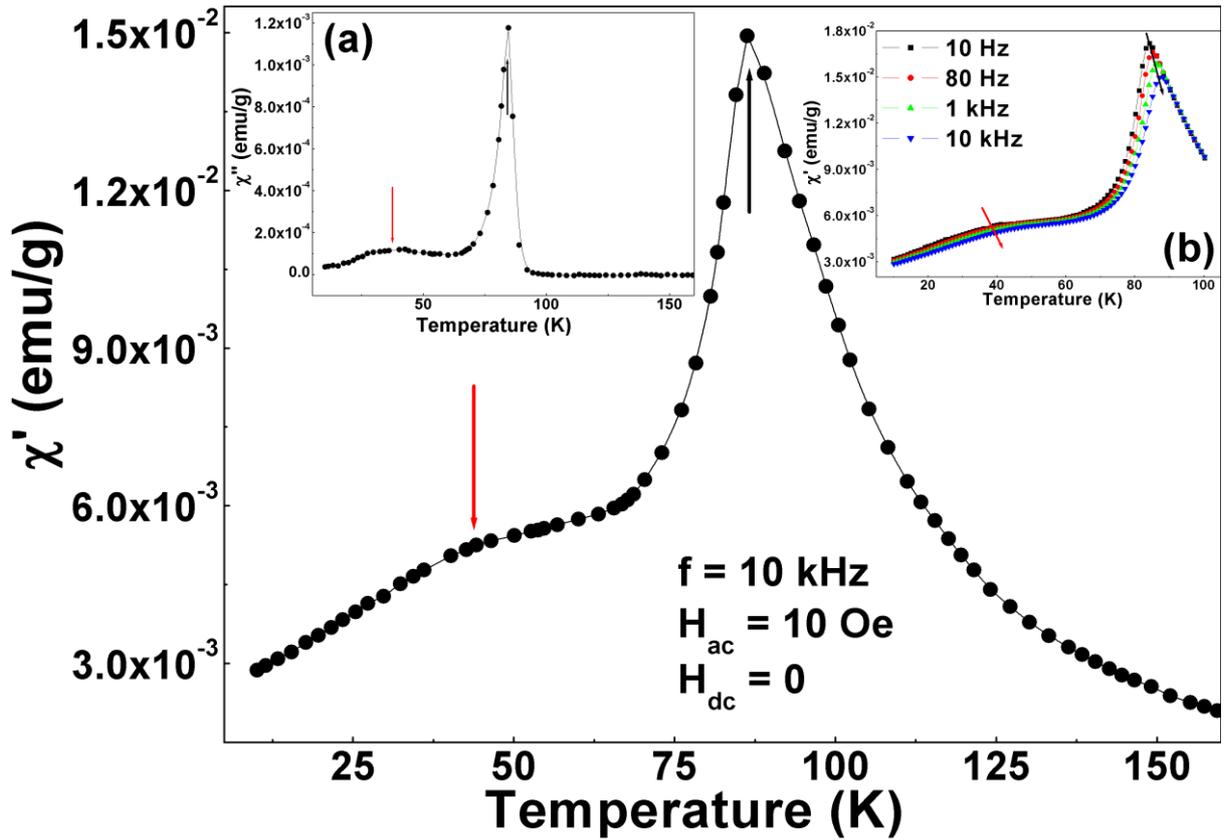

Fig. 7. Temperature dependence of the real (in-phase) component of a. c. susceptibility for YBaFe$_{3.6}$Ga$_{0.4}$O$_7$ as a function of temperature measured in zero magnetic field (H$_{dc}$ = 0), using a frequency of 10 kHz. Inset (a) shows the imaginary (out-of-phase) component of the a. c. susceptibility, and inset (b) shows the real (in-phase) component of the a. c. susceptibility measured using four different frequencies.

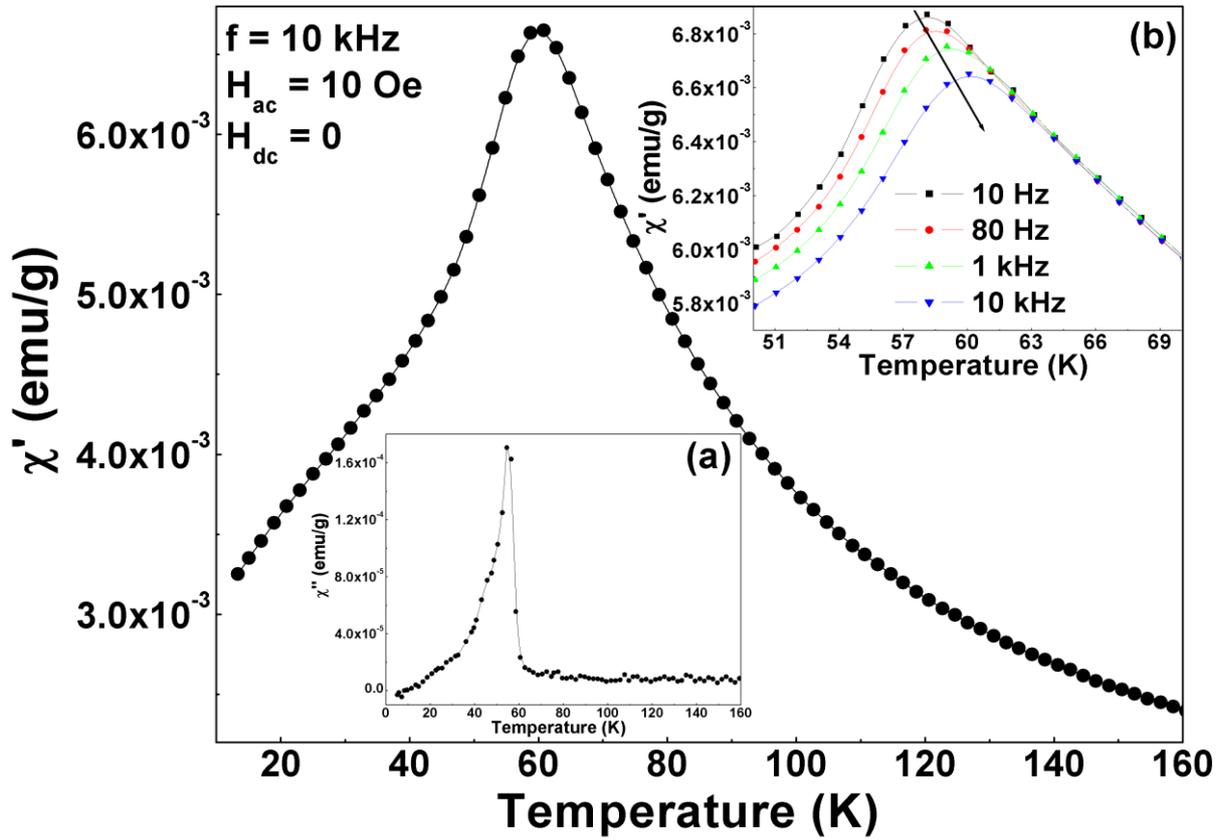

Fig. 8. Temperature dependence of the real (in-phase) component of a. c. susceptibility for YBaFe$_{3.3}$Ga$_{0.7}$O$_7$ as a function of temperature measured in zero magnetic field (H$_{dc}$ = 0), using a frequency of 10 kHz. Inset (a) shows the imaginary (out-of-phase) component of the a. c. susceptibility, and inset (b) shows the real (in-phase) component of the a. c. susceptibility measured using four different frequencies.

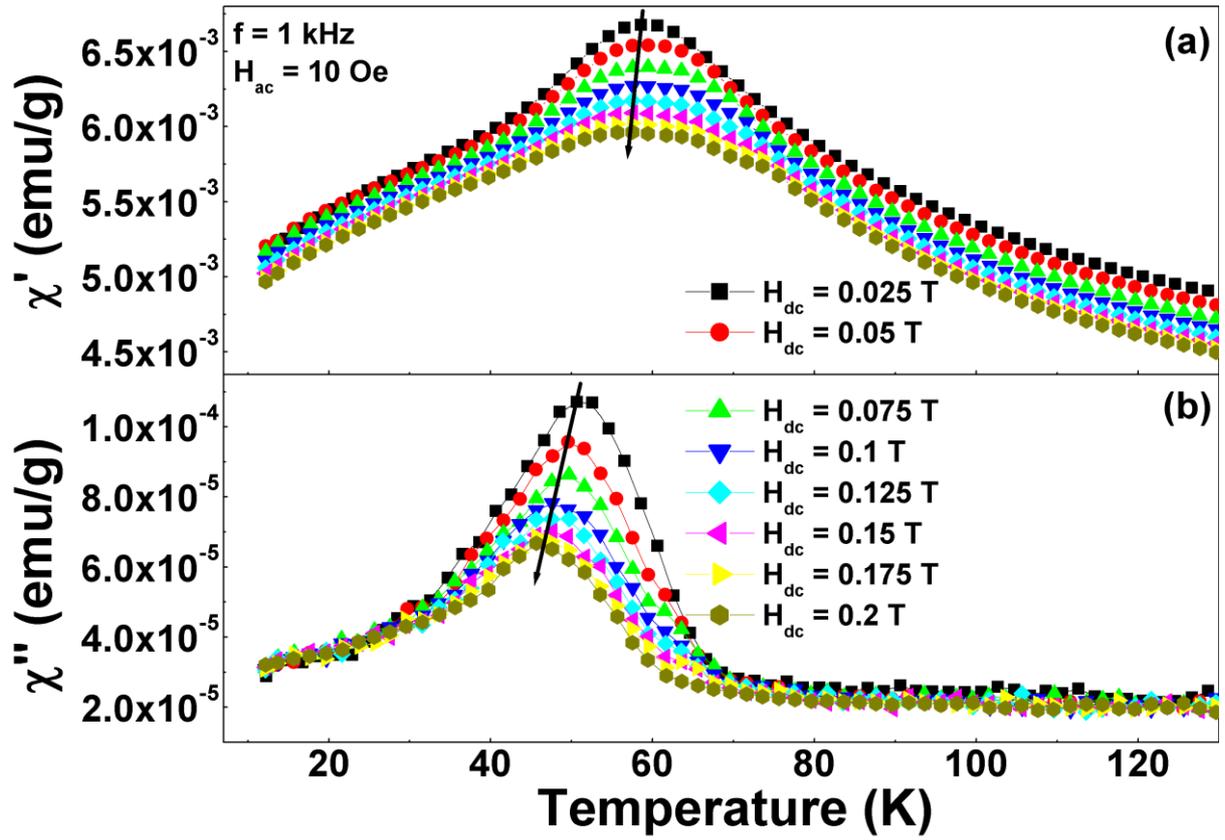

Fig. 9. The (a) real (in-phase) and (b) imaginary (out-of-phase) component of a. c. susceptibility for YBaFe$_{3.3}$Ga$_{0.7}$O$_7$ as a function of temperature. The driving frequency was fixed at $f$ = 1 kHz and H$_{ac}$ = 10 Oe. Each curve was obtained under different applied static magnetic field (H$_{dc}$) ranging from 0 T to 0.2 T.

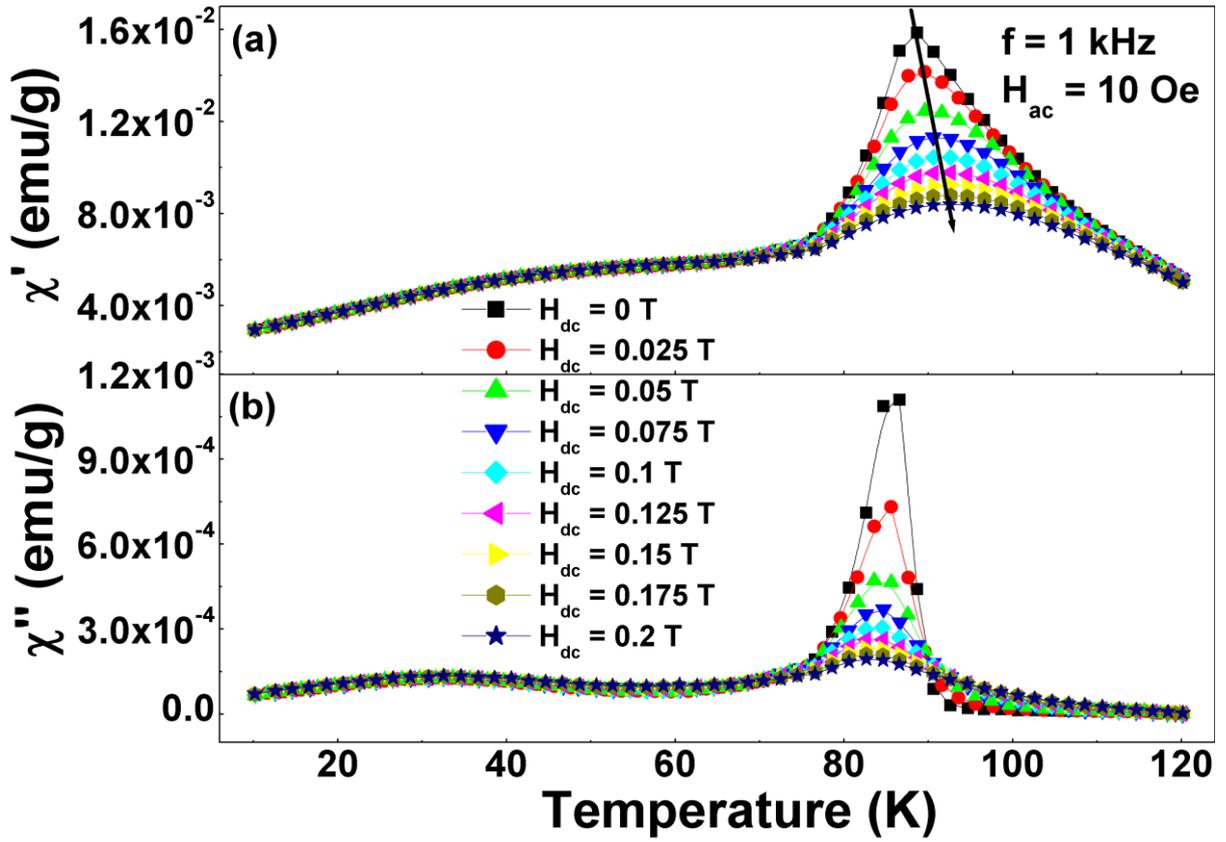

Fig. 10. The (a) real (in-phase) and (b) imaginary (out-of-phase) component of a. c. susceptibility for YBaFe$_{3.6}$Ga$_{0.4}$O$_7$ as a function of temperature. The driving frequency was fixed at $f$ = 1 kHz and H$_{ac}$ = 10 Oe. Each curve was obtained under different applied static magnetic field (H$_{dc}$) ranging from 0 T to 0.2 T.